# Deep Learning-Based Brain Image Segmentation for Automated Tumour Detection

Suman Sourabh[1], Murugappan Valliappan[1], .Narayana Darapaneni[2] and Anwesh R P[3,*]

[1] PES University, Bangalore, Karnataka, 560050, India
[2] Northwestern University, Evanston, IL 60208, United States
[3] Great Learning, Hyderabad, Telangana, 500089, India

## Abstract

INTRODUCTION: The present study on the development and evaluation of an automated brain tumor segmentation technique based on deep learning using the 3D U-'Net' model.
OBJECTIVES: The objective is to leverage state-of-the-art convolutional neural networks (CNNs) on a large dataset of brain MRI scans for segmentation.
METHODS: The proposed methodology applies pre-processing techniques for enhanced performance and generalizability.
RESULTS: Extensive validation on an independent dataset confirms the model's robustness and potential for integration into clinical workflows. The study emphasizes the importance of data pre-processing and explores various hyper parameters to optimize the model's performance. The 3D U-Net, has given IoUs for training and validation dataset have been 0.8181 and 0.66 respectively
CONCLUSION: Ultimately, this comprehensive framework showcases the efficacy of deep learning in automating brain tumour detection, offering valuable support in clinical practice.

## 1. Introduction

Deep learning and artificial intelligence have made considerable strides recently, especially in medical picture processing. New techniques that make it simpler and faster for doctors to diagnose diseases have been developed because of this. To avoid life-threatening diseases and disorders, the biomedical area needs accurate and effective information, which can only be provided by computer-assisted technology. The lack of annotated data is one of the key obstacles to accelerating the development of deep-learning in clinical or medical applications. Recent developments in deep learning algorithms, however, have demonstrated that larger datasets produce greater prediction performance and higher accuracy. A dangerous and unpredictably developing kind of cancer in brain known as a brain tumor can seriously harm the brain.

New techniques for identifying brain tumors from medical photos have been developed thanks to recent developments in deep learning. This represents a significant advance since it may enable earlier recognition and management of brain cancers, potentially improving patient outcomes.

In our research, we have developed a revolutionary deep-learning method for identifying brain cancers/tumors. Our approach is based on a convolutional neural network-CNN called the U-'Net' design architecture, which was created especially for segmenting medical images. We trained our model on a dataset of brain MRI scans from BraTS'2020 and tested it on a batch of unseen images, Anatomical predictions made by our model have been 99.13% accurate.

Here are some additional details about brain-tumors:

- Benign and malignant brain-tumors are the two main varieties. Benign tumors do not spread to other parts of the body and has no cancerous symptom. Malignant

- cancerous tumors have the capacity to spread to other organs.
- A brain tumor may occur anywhere in the brain. The three most typical types of brain tumors are meningiomas, gliomas and pituitary tumors.
- The size and location of the tumor can affect the symptoms of a brain tumor. Headaches, nausea, vomiting, seizures, and visual issues are a few of the typical symptoms.
- Brain tumors are often diagnosed with a combination of MRI, CT scan, and biopsy.
- Based on the size, kind, and location of a brain tumor different course of treatment applied are Chemotherapy, radiation therapy, and surgery.

Brain tumors are a serious condition, but early diagnosis and treatment can improve patient outcomes. Deep learning methods like the one we have developed have the potential to help doctors diagnose brain tumors earlier and more accurately, which could lead to better patient outcomes.

## 2. Related Works

There were 3 different most common types of brain malignancies, including meningiomas, gliomas and pituitary tumors, were classified by Chelghoum et al., 2020 [1] using the brain contrast-enhanced MRI benchmark dataset. Reduce training time, get rid of overfitting, and improve classification accuracy were their goals. Transfer learning was used to classify brain tumors using 9 pre trained deep-networks, including ResNet101, GoogleNet, ResNet-Inception-v2, VGG16, SENet, VGG19, ResNet18 , ResNet50, & AlexNet. A classification accuracy of 98.71% was attained.

A concept for categorizing brain MRI images into carcinogenic and non-cancerous was presented by Khan et al., 2020 [2] by combining data augmentation and image processing with a CNN - convolutional neural network. They proposed a novel method for classifying brain cancers that first uses image edge recognition to identify the area that is relevant in MRI images, then crops it. To expand the amount of training data, they also used data augmentation. They discovered that an effective tool for classifying brain tumors was a straightforward CNN network.

According to Sangeetha et al., 2020 [3], a number of CNN-based classification techniques, such as ResNet, VggNet, and GoogleNet were recommended. They discovered that ResNet50 produced the greatest outcomes, with 10% more accuracy in 10% less time than VggNet and GoogleNet.

A method for categorizing various brain modalities and brain-tumors was proposed by Sharif et al., 2019 [4]. Their process comprises two stages:

- The Dice Similarity Rate (DSR) is used to segment the tumor using the SbDL approach.
- PSO – Particle-swarm-optimization  is used to improve the deep learning fusion and characteristics of the discriminative region-based local binary pattern (DRLBP), which are then utilized to identify the various tumor classifications using a softmax classifier.

In a survey conducted by Muhammad et al., 2020 [5], they analyzed the literature already in existence and the most recent deep learning techniques for classifying brain tumors. The study's primary focuses were feature extraction, pre-processing, and the categorizing of brain tumors as a whole. It looked at the successes and shortcomings of these strategies and investigated sophisticated CNN - convolutional neural network representations for classifying brain tumors. The authors tested transfer learning in-depth both with and without data augmentation.

For the classification of brain cancers, Saleh et al., 2020 [6] has used 5 pre trained design architectures: Xception, MobileNet, ResNet50, InceptionV3, and VGG16. Their framework's major goal was to improve the ability of MRI scanning technology to categorize various brain tumor kinds and locate certain tumor subtypes. The best accuracy rate of 98.75% was attained by the Xception architecture, which may help in the early diagnosis of a variety of brain cancers.

The viability of applying deep learning and transfer learning frameworks to segment MRI images precisely, detecting, and analyzing LGG (low-grade glioma) brain tumors was established by Naser et al., 2020 [7]. To accomplish fully automated and simultaneous segmentation of MRI brain tumors, they created a pipeline that combines segmentation and grading frameworks. This strategy guarantees a thorough automation of the procedure.

A model that uses unique components to identify specific areas on MRI images and differentiate between healthy and unhealthy brain MRI images was proposed by Togacar et al., 2019 [8]. 96.15% of the classification attempts were successful.

A unique and effective convolutional neural network (CNN) known as RWNN-Randomly wired neural network was introduced by Xiaohao et al., 2020 [9]. The RWNN's structural relations were constructed at random. According to experimental findings, the augmented technique increased test accuracy by about 1% over the baseline RWNN model. When it came to segmenting images of brain tumors, the improved RWNN model performed comparably to earlier studies and other convolutional neural network models like EfficientNet and ResNet.

A reinforcement learning method was developed by Cicek et al., 2016 [10] to partially and totally automatically partition a 3D dimension from a brief description. The Xenopus kidney's extraordinarily complex components, which were organized and led from root to tip, were precisely segmented using the framework.

Transfer learning was suggested by Polat et al., 2021 [11] as a method for classifying various types of brain-tumors from T1 MRI images. Pre-trained weights from the VGG16, DenseNet121, Res-Net-50, and VGG-19 architectures were used in the transfer learning method. Transfer learning is a time- and resource-saving technique because it makes use of the knowledge learned through training networks on huge datasets. utilizing the Adadelta optimization method, the total classification performance exceeded other methods utilizing the same dataset, with DenseNet121 and ResNet50 reaching classification efficiencies of 98.92% and 99.20%, respectively.

With an emphasis on MRI-based neuro medical imaging, Al Galal et al., 2021 [12] carried out systematic deep learning research. A Deep-learning technique with Convolutional neural network have been found to be really useful in several biological imaging subfields, including standardization, recognition, and categorization. The authors used methods including data augmentation & transfer learning to overcome issues with the deep learning approach, such as little information and tags.

An improved version of the U-'Net' known as the "DIU-'Net'," which incorporated features from DenseNet and GoogleNet, was proposed by Zhang et al., 2020 [13]. The Inception Res block, Dense Inception block, down sampling block, and up inspecting block were the 4 key components of the DIU-'Net'. The experimental findings suggested, the DIU-'Net' model executed better than methods like U-'Net', and ResU-'Net', SegNet, FCN-8s used previously.

In their model, Isensee et al., 2020 [14] used nnU-'Net' to BraTS'2020 challenge. Results from the initial nnU-'Net' configuration were already satisfactory. However, integrating BraTS-specific modifications, such as post-processing techniques, region-based training, more aggressive data-augmentation, and minor tweaks to the U-'Net' channel, considerably improved the segmentation performance.

Transfer learning was used by Kaur et al., 2020 [15] to assess various pre trained Deep-convolutional neural network models for segmenting MRI images. A high level of recognition rate was attained employing the method of using transfer learning and pre-trained DCNN models. The AlexNet model outperformed all other examined models, earning classification accuracy rates of 100%, 94%, and 95.92% for the three datasets.

On the BraTS'2020 dataset, Maram and Rana, et al., 2021 [16] employed the U-'Net' model for segmentation. They scored 98.48% accuracy.

## 3. BRATS Dataset Specification

The BraTS'2020 image segmentation dataset is a collection of MRI brain scans. All BraTS multimodal scans are available as NIfTI standard files (.nii or . nii.gz)-Neuroimaging Informatics Technology Initiative. It is an open file format for storage of medical image data (historically used for neuroimaging – hence the name, but not restricted to neuroimaging). NIfTI files are used to efficiently store medical data together with additional necessary metadata. It is mainly used in research and development setting as an example to develop an application. NIfTI file structure has Header that contains information mainly about image geometry (resolution, position, orientation) and Body that contains actual image pixel data (2D, 3D, 4D). The file extension usually ".nii" or ".nii.gz" (compressed version). NiBabel is the library in python to deal with the NIfTI files. NIfTI header only contains relevant information about the image or of the volume, such as the affine matrix and shape, no information about the patient is kept.

The dataset consists of 4MRI modalities [Fig. 1]:

- T1: Its slice has thickness of 1 mm to 6 mm, T1-weighted native-images have been acquired in axial 2D.
- T1ce: This image was taken in three dimensions, is T1-weighted, and was contrast-enhanced with gadolinium for the majority of the patients.
- T2: Axial 2D capture of a T2-weighted image and it has a slice thickness of 2 to 6 mm.
- FLAIR: 2–6 mm slice thickness, axial, coronal, or sagittal 2D captures of T2-weighted FLAIR images. Data were collected from numerous (n=19) institutes using diverse scanners and clinical regimens.

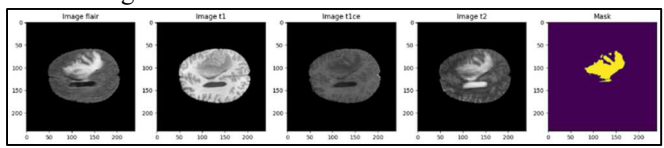

Figure. 1. Example of Sample data available in BraTS2020 Dataset.

Each MRI dataset was individually segmented and approved by an expert neuroradiologist.
Annotations (labels) are provided as described in Table-1:

Table-1: Labels available in BraTS'2020 Dataset

| Labels | Description |
|---|---|
| 0 | Unlabeled Volume |
| 1 | NCR/NET- Core tumor that is necrotic and not increasing |
| 2 | ED- Peritumoral-edema |
| 3 | Missing in all segmentation images. |
| 4 | GD- enhancing Tumor (ET) |

## 4. Proposed Method

Our approach steps [Fig 3] in this paper are as below:
- Get the data ready.
- Define custom data generator.
- 3D U-'Net' model
- Train and Predict

All the steps shown in the fig. 3 are explained the next sub sections IV-A to IV-D.

### 4.1 Get the data ready

The BraTS'2020 dataset is available in 4 volumes(T1, T2, T1CE, and FLAIR). The native volume T1 does not include a lot of information for segmentation, therefore after checking and comparing these volumes with the corresponding mask, we have eliminated it(T1) and combined the 3 non-native volumes (T2, T1CE, and flair) into a single multi-channel volume. Fig. 1 shows that the images t1ce, t2, and flair provide the most information on the mask. Thus, in order to limit the amount of data and computation required, native image t1 is not taken into account for training.

As described in Table 1, in all the mask present in the Training set the label 3 is missing hence we have reassigned the pixels of value 4 to 3. Now the segments have the label aligned with [0.,1.,2.,3.]. also, the labels are present with datatype float, we have changed it to unassigned int (unit8) [class- 0, 1, 2, 3] so that it will consume less space in the memory.

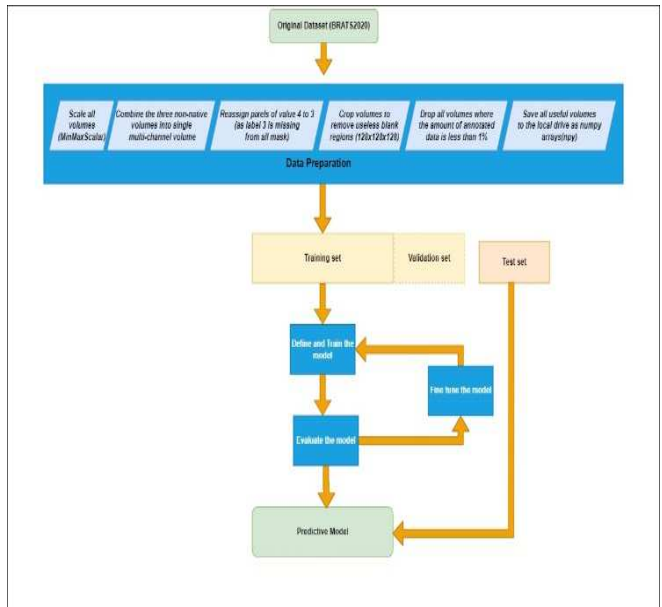

Figure. 2: proposed method design flow

The original file has size (240x240x155). For the volumes T1CE, T2 and FLAIR (T1 is not in consideration to train the model). We have eliminated pointless blank areas surrounding the actual volume of interest by cropping the image to size (128x128x128).

The pixels assigned in the data have been normalize with MinMaxScalar (min=0, max=1).

Additionally, we have removed all volumes whose annotated data content fell below a certain threshold (1%). After making all these adjustments to the training data, we have divided the image and mask volumes into training and validation datasets and stored all important volumes to the local storage as NumPy arrays (npy).

## 4.2 Custom data generator

As we have saved the training data as NumPy array(.npy files) in our design approach, hence we have created a custom data generator to feed the files from the local disk because the Keras default generators only works for .jpg, .png and .tif files.

## 4.3. 3D U-'Net' model

In our study, semantic segmentation has been carried out using a 3D U-'Net' design model [fig. 3]. In order to improve feature extraction and fusion, the model has been built employing several number of convolutional and pooling layers, as well as skip connections. The contraction and the expansion path are the two main paths in the model.

**Input Layer:**

The input layer is defined with the shape of the input data. After reducing the size of the input data as described in section IV-A, we have taken (height=128, width=128, depth=128, channel=3) as a size of input.

**Convolutional Layers:**

3D convolutional layers are applied one after the other in succession. Each convolutional layer in this code is defined using Conv3D.

As we delve further into the network, the number of filters or feature maps grows. With each convolutional layer, the number of filters in this algorithm has been increased by two, starting at 32. 32, 64, 128, 256, and 512 filters have been used.

The initialization method 'he_normal', which aids in avoiding the vanishing gradient issue during training, has been set as the kernel_initializer.

Rectified Linear Unit, the activation function utilized, introduces nonlinearity into the model.

**Dropout Layers:**

A dropout rate of 0.1 or 0.2 has been added in drop out layer after every convolutional layer, which helps preventing overfitting problem.

**Pool Layers:**

A 3D max-pooling layer (2, 2, 2) has been added after each pair of convolutional layers to shrink the feature maps spatial dimensions.

**Progression:**

To build a deep network, a series of convolutional, dropout, and max-pooling layers have been repeated. Due to max pooling, the spatial dimensions get smaller with each repetition while the quantity of filters used to capture higher-level information grows.

The output of the final convolutional-layer has been used to upsample the feature maps back to their original spatial dimensions using the Expansive Path (Decoder or Upsampling Path).

The expansion path: From the features that the contraction path extracted, the expansion path is in charge of recreating the image. A concatenation layer has been added after a few transposed convolutional layers, often referred to as upsampling or deconvolution, have been merged with skip connections from the corresponding layers in the contraction path. The number of channels or filters and the size of the pool window have been reduced with each iteration of the process. As a result, the image can be recreated by the model in its original resolution.

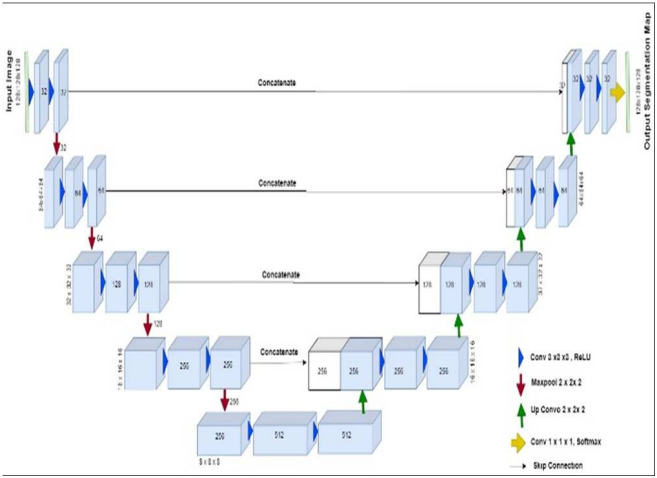

Figure. 3: Proposed 3D U-'Net' Model

**Up sampling Layers:**

Upsampling layers - deconvolutional layers or Transposed convolutional layers, are the first layers in the Expansive Path. Conv3DTranspose has been used to specify each transposed convolutional layer in this model.

Each transposed convolutional layer's strides argument has been set to (2, 2, 2), upsampling the feature maps by a factor of 2 in each dimension. Each transposed convolutional layer resulted in a doubling of the spatial dimensions.

**Concatenation (Skip Connections):**

Using the concatenate function from Keras, the output feature maps have been joined with the corresponding feature maps from the Contraction Path after each transposed convolutional layer.

**Convolutional Layers:**

The concatenated feature maps are then subjected to a number of 3D convolutional layers after the concatenation phase. Each convolutional layer in the Expansive Path features 3x3x3 kernels, ReLU-activation function, and 'he_normal' kernel initializer, just like the Contraction Path.

Convolutional layers have the same number of filters as the Contraction Path, but because the feature maps have been concatenated, there are twice as many channels as before.

**Dropout Layers:**

For our proposed model we have selected dropout rate of 0.1 or 0.2 (same as in the Contraction Path) for regularization after each convolutional layer in the Expansive Path.

**Progression:**

To build a deep network in the Expansive Path, a series of transposed convolutional, concatenation, convolution, and dropout layers are repeated. Each repetition doubles the number of channels, upsamples the feature maps, and recovers spatial information via skip connections.

**Output Layer:**

A 1x1x1 convolutional layer with the softmax activation function concludes the Expansive Path. The final segmentation masks for each class are created using the output of the convolutional layer.

The compilation has been carried out outside of this function to allow for the use of various optimizers, loss functions, and metrics.

For semantic segmentation, this code constructs a 3D U-'Net' model using convolutional and transposed convolutional layers. It can be applied to projects like the segmentation of medical images or other 3D picture jobs with many classes.

**Training and Prediction:**

The proposed model has been trained with Learning Rate = 0.0001, Optimizer = Adam, epoch =100 and Training accuracy of 99.14% has been achieved whereas the Validation accuracy was 98.18%.

The proposed model is compiled with the below parameters:

Metrics: The metrics are employed to assess the model's performance during both training and testing. The metrics has been set to ['accuracy', sm.metrics.IOUScore(threshold=0.5)]. This indicates that the accuracy measure and the IoU - Intersection over Union metric has been be used to assess the model. The proportion of pixels that are correctly categorized is measured by the accuracy metric. The overlap between the predicted segmentation and the actual segmentation has been measured by the Intersection over Union-IoU metric.

Optimizer: During training, the optimizer is responsible for changing the model's weights. The optimizer in this situation is set to keras.optimizers.Adam(), a well-liked deep learning model optimizer. The learning rate for the optimizer is represented by the LR argument, which is set to 0.0001.

Loss Function: The model's performance during training has been evaluated using the loss function. We have used both dice loss and focal loss to our model with formula (3). For image segmentation tasks, the dice loss is a loss function that is frequently utilized. The problem of class imbalance has been intended to be solved by the loss function known as the focal loss.

Dice drop: A common loss function for picture segmentation task is dice loss (1).

$$DiceLoss(Y\_true, Y\_pred) = 1 - \frac{\sum(2 * (Y\_true * Y\_pred)}{\sum Y\_true + \sum Y\_pred}$$

(1)

The loss function has the following characteristics:

- It is robust, so it is less sensitive to small changes in the predicted mask.
- It is differentiable, so it can be used with gradient descent optimization algorithms.

- It is symmetric, so it does not penalize one class more than another.

Focal loss: Deep learning tasks involving object detection use a loss function called focal loss (2). It is intended to deal with the class imbalance issue that frequently arises in datasets for object detection. The class imbalance issue arises when there are comparatively large number of background and few foreground examples in the collection. As a result, learning to accurately categorize the foreground instances may be challenging for the model.

$$FocalLoss(P_t) = -(1 - P_t)^\gamma * log(P_t) \quad (2)$$

Where:
$P_t$ : Probability of model correctly classifying the example
γ: Hyperparameter that controls the degree of focus.

Dice and focal loss have been combined to form 'total loss' to calculate the loss in proposed model training.

Total_loss = DiceLoss + (1 * FocalLoss)   (3)

The model can segment tumors with high accuracy, and it can be used to help doctors diagnose and treat brain tumors.

## 5. Results and Discussion

The model underwent 100 training epochs, with 129 iterations on training photos in each epoch. The complete training has taken 36 hours on the complete dataset. On BraTS'2020 dataset, training and validation accuracy for the 3D U-'Net' design architecture have been achieved as 99.13% and 98.18%, (Fig. 4- accuracy graph). The total losses for training and validation were 0.7803 and 0.8147, respectively (Fig. 5- ). The IoUs for training and validation dataset have been 0.8181 and 0.66 respectively (Fig. 6). The Google-Colab environment (Jupyter notebook) has been used to develop the entire framework, while Google Drive was used to mount the 40 GB input dataset. There may be plenty of room to fine-tune these hyper parameters and enhance the model's performance.

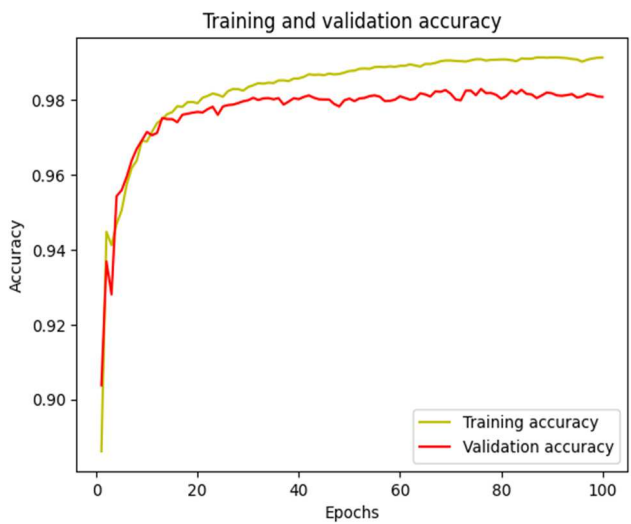

Figure. 4: Accuracy

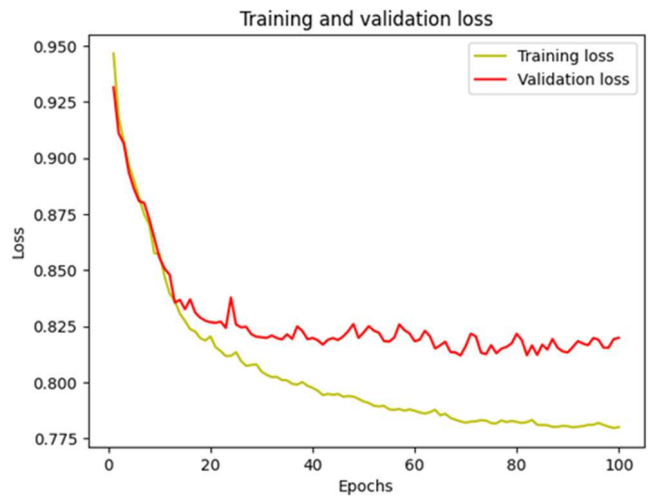

Figure. 5: Total-loss

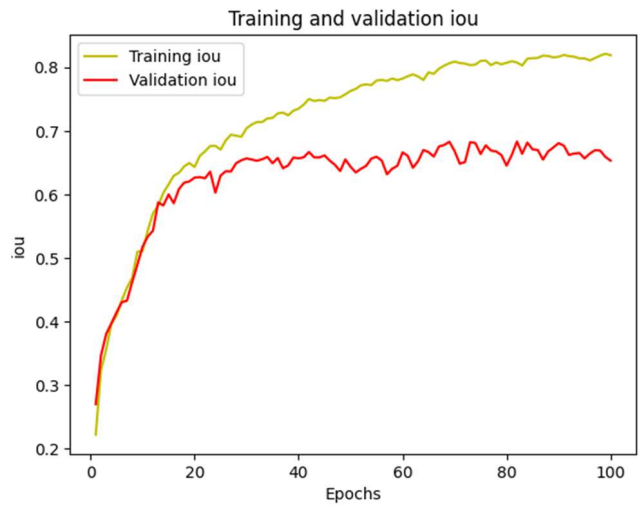

Figure. 6: IoU

The segmentation prediction on the random test image by the model has been shown in Fig 7.

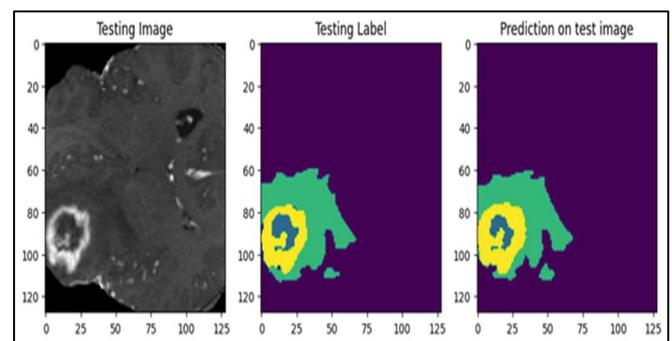

Figure. 7: Test image generation on the proposed 3D-U-'Net' Model

## 6. Comparative Study

Our method has achieved a higher accuracy of 99.13% than the base U-'Net' method [16] with an accuracy of 98.45%. This shows that the proposed 3D U-'Net' model, is better suited for segmenting brain-tumors in MRI images than a 2D U-'Net' model.

In addition, the proposed method uses a deeper network architecture than the U-'Net' method. This allows the proposed method to learn more complex features from the MRI images, which also contributes to its higher accuracy.

Overall, the proposed method is the more accurate method for brain tumor segmentation on the BraTS'2020 dataset. It achieves an accuracy of 99.13%, which is significantly higher than the U-'Net' method[16].

Table 2: Comparison with 2D U-'Net' research article

| SOURCE | DATASET | TYPE OF IMAGE | USED METHOD | ACCURACY |
|---|---|---|---|---|
| [16] | BraTS'2020 | MRI | U-'Net' | 98.458% |
| Proposed Method | BraTS'2020 | MRI | 3D U-'Net' | 99.13% |

## 7. Future Scope and Conclusion

In particular, the U-'Net' design architecture used in the MICCAI BraTS'2020 dataset is the subject of this work. In comparison to alternative architectures, the trained 3D U-'Net' model had training and validation accuracy of 99.13% and 98.18%, which are thought to be better results.

On the BraTS'2020, the Residual Network can be used in conjunction with the U-'Net' model to further increase accuracy. Accuracy has been increased because of the combination strategy. Additionally, the 3D U-'Net' design fared better than the conventional U-'Net' framework at approximating tumors in 3-dimensional biological pictures. The 3D U-'Net' model's hyperparameter tuning can produce results that are more accurate.

Data pre-processing is essential for medical image analysis and includes methods like data augmentation. By removing unnecessary data and information from the images, these pre-processing techniques help to improve analysis and interpretation.